\documentclass[twocolumn,aps,floatfix,showpacs,superscriptaddress]{revtex4}

\usepackage{amsmath,amssymb,eucal,graphicx}
\usepackage{color}
\usepackage{ulem}

\newcommand{\p}[1]{{}}

\begin{document}

\title{Asymmetric Exclusion Process with Global Hopping}

\author{Chikashi Arita}
\affiliation{Institut de Physique Th\'eorique, IPhT, CEA Saclay
and URA 2306, CNRS, 91191 Gif-sur-Yvette cedex, France}
\affiliation{Theoretische Physik, Universit\"at
 des Saarlandes, 66041 Saarbr\"ucken, Germany}
\author{J\'er\'emie Bouttier}
\affiliation{Institut de Physique Th\'eorique, IPhT, CEA Saclay
and URA 2306, CNRS, 91191 Gif-sur-Yvette cedex, France}
\affiliation{D\'epartement de math\'ematiques et applications, \'Ecole
normale sup\'erieure, 75230 Paris cedex 05, France}
\author{P. L. Krapivsky}
\affiliation{Institut de Physique Th\'eorique, IPhT, CEA Saclay
and URA 2306, CNRS, 91191 Gif-sur-Yvette cedex, France}
\affiliation{
Department of Physics, Boston University, Boston, MA 02215, USA}
\author{Kirone Mallick}
\affiliation{Institut de Physique Th\'eorique, IPhT, CEA Saclay
and URA 2306, CNRS, 91191 Gif-sur-Yvette cedex, France}

\begin{abstract}
We study a one-dimensional totally asymmetric simple exclusion process
with one special site from which particles {\it fly} to {\it any}
empty site (not just to the neighboring site). The system attains a
non-trivial stationary state with a density profile varying over the
spatial extent of the system. The density profile undergoes a non-equilibrium phase transition
when the average density passes through the critical value $1-[4(1-\ln 2)]^{-1}=0.185277\ldots$,
viz. in addition to the discontinuity in
the vicinity of the special site, a shock wave is formed in the bulk
of the system when the density exceeds the critical density.
\end{abstract}


\maketitle

\section{Introduction}

The asymmetric simple exclusion process (ASEP) is a well-studied model
of low-dimensional transport of particles with hard-core
interactions. This model has become a standard tool in the context of
low-dimensional transport, and it is commonly used to represent the motion
of molecular motors or more generally enzymes along cytoskeletal
fibers. It is interesting to recall that the ASEP was originally
proposed in 1968 as a model of the kinetics of biopolymerisation in
RNA templates \cite{MGP}. Since then, the ASEP has been
extensively studied and it has achieved the status of a paradigm in
statistical physics (see \cite{Sp,K,HZ,Sch1,BA,D1} and references
therein). It is fair to say that the ASEP is one of the simplest
non-trivial models of a non-equilibrium process: it is an interacting
$N$-body system without detailed balance, which reaches a non-Gibbsian
stationary state with non-vanishing currents. In the last forty years,
exact solutions for the ASEP in various contexts have been obtained
thanks to increasingly sophisticated and elegant mathematical
methods \cite{DEHP,SD,DEMu,DEMa,Sch2,PM,TW,D2}. 
Conversely, these theoretical studies together with
experimental progresses in micro-manipulations have triggered a
renewed interest in biophysical applications of the ASEP in recent
years  \cite{GCCR,CSN,TPPRC,CSR,TEK,CLBCJ,JEK,MRF,EAS}
(see \cite{CKZ} for a review).

The presence of a non-vanishing current in a stationary state results in 
the transport of information from one part of a system to another, leading to
long-range correlations. Henceforth, the behavior of the
ASEP is highly sensitive to boundary couplings or to the presence of
local defects. In particular, the ASEP can undergo boundary induced
transitions even in one dimension \cite{DEHP}, which can not occur in
one-dimensional equilibrium systems with short range
interactions. Hence, simple deformations of the ASEP are often very
challenging. These models are important since they enable us to gain  
insight into  macroscopic and microscopic behaviors. One such
seemingly innocent generalization is to introduce blockages. This is easy to realize for 
the ASEP on a periodic one-dimensional lattice where particles undergo
a biased nearest-neighbor hopping to empty sites; if the ring is
homogeneous a stationary state with all permissible configuration
being equiprobable is reached (on the macroscopic level, the density
is uniform in the stationary state).  Suppose that the hopping rate
between one pair of neighboring sites is suppressed by a certain
factor.  This inhomogeneous model with two parameters (the suppression
factor and the average density)  is a well-known challenge that has
not been exactly solved \cite{JL1,Gunter,JL2}; see \cite{CLST} for a
review and recent progress. The chief message is clear: A local change
of the environment (one special blockage bond) can result in a global
change, namely the system segregates into high- and low-density
phases. However, details of the density profiles as well as the
stationary-state correlations are still unknown \cite{CLST}.

In the present work, we consider a variant of the ASEP
on a one-dimensional lattice, namely on a ring with $L$ sites. The total number of particles $N$ is conserved by dynamics. The system is assumed to be homogeneous apart from one special site from which the particle can {\it fly} to any other site. Hence in contrast to Refs.~\cite{JL1,Gunter,JL2} we investigate the effect of a special site 
rather than a special bond. Our basic motivation  is to define a simple model with {\it unique} parameter, the average density $\rho = N/L$ (or equivalently the
fraction of the vacant sites $v=1-N/L$),  which can exhibit different
regimes with respect to the different values of  $\rho$. We show that
introducing one special site with global hopping generates a non-trivial
phase diagram and leads to a subtle selection mechanism of the
stationary density profile.

The outline of this work is as follows. In Sec. \ref{sec:basic}, we
explain the dynamical rules of our model. In Sec. \ref{sec:hydro},
we write down the governing equations characterizing large scale
stationary hydrodynamic behavior. In Secs.~\ref{sec:smooth} and \ref{sec:shock}, 
we investigate two types of density profiles that can
occur,  and we show that naive reasoning does not provide a suitable
selection mechanism for the mean-field equations. A more precise
analysis (Sec.~\ref{sec:selection}) allows us to establish the phase
diagram of the system which,
despite the simplicity of the model, turns out to be rather rich. Concluding
remarks are given in Sec.~\ref{sec:concluding-remarks}.

\section{The Model}
\label{sec:basic}

\begin{figure}
 \includegraphics[width=10cm]{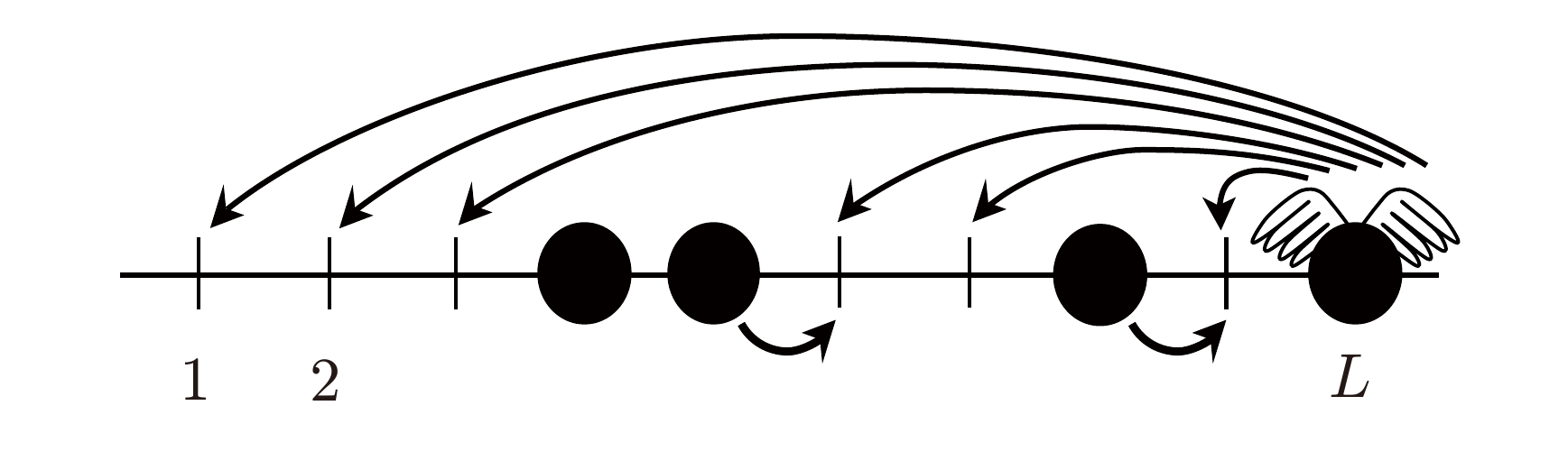} 
\vspace{-5mm}
\caption{The TASEP with {\it global hopping} from one site.}
\label{fig:wing}
\end{figure}

We consider the totally asymmetric simple exclusion process (TASEP)
on a one-dimensional   lattice  with $L$ sites, which are labeled by
$j=1,\ldots,L$. Each site  $j$ is either occupied by one particle
($\tau_j=1$)  or empty ($\tau_j=0$).  In an infinitesimal time
interval $dt$, a particle on site $j$ ($1 \le j \le L-1$) attempts to
hop forward to site $j+1$ with probability $dt$; the attempt is
successful only when the target site is empty. A particle occupying
the rightmost site $L$ can  {\it fly} to any site $j=1,\ldots,L-1$; a
flying attempt is made with probability $dt$ and is allowed if the
target site (chosen uniformly at random) is empty, see Fig.
\ref{fig:wing}. These rules define our basic model. The total number
of particles $N$ is a conserved quantity. The only  free parameter in
this model is the density $\rho = N/L$, or equivalently the fraction $v = 1 -\rho$ of vacant sites. 

Monte Carlo simulations show that the system attains an interesting
stationary state which exhibits a qualitative change when the average
density passes through the critical value; we will show below that the critical
density equals  $1-[4(1-\ln
  2)]^{-1}=0.185277\ldots$.  Our approach can be extended to a
generalized system in which the special site is unique not only since
the hopping from this site is global,  but also the hopping rate is
arbitrary. More precisely, we postulate that in an infinitesimal time
interval $dt$, a particle at the special site $L$ attempts to fly with
rate $\beta dt$, where $\beta$ is a positive real number, and the
target site is chosen uniformly at random amongst the $vL$ {\it empty sites.} 
We shall derive the phase diagram of the full model in the $(v,
\beta)$-plane.  The original model corresponds to the case $\beta=v$.
The goal of this article is to characterize analytically the phase
diagram of the model.
We note that, by regarding holes as particles and particles as holes, the process we study is a
many-particle generalization of the  stochastic-resetting problem \cite{EM}.

\section{Large Scale Hydrodynamic Behavior}
\label{sec:hydro}

A macroscopic (or hydrodynamic) description is valid when the average density slowly varies.
In such situations, instead of the density per each site $\rho_j=\langle \tau_j \rangle $, 
one can study
 the quantity $F(x)$
  which depends on the macroscopic variable $x=j/L$. 
The macroscopic density can be defined as 
\begin{equation}
F(x) = \frac{1}{2\ell + 1}\sum_{k=j-\ell}^{j+\ell} \rho_k
\label{HydrProf}
\end{equation}
where $1\ll \ell \ll L$. We study $F(x)$ by employing a hydrodynamic
approach, mostly the Eulerian inviscid treatment. Such an approach
cannot provide a fair description of the regions inside shock waves
and boundary layers, but it can be trusted away from such narrow
regions. In the stationary state,  we have
\begin{equation}
\label{Euler}
\frac{d}{dx}\,F(1-F) =  \frac{\beta}{v}\, \rho_L(1-F).
\end{equation} 
The average density $\rho_L$ at the special point $j=L$ requires a
special treatment:  A boundary layer  could appear near this point, and the
limiting hydrodynamic density (we denote it by $R$) may be  different from
the true microscopic density, i.e., we have  generically 
\begin{equation}
\label{lim_right_density}
\lim_{x\to 1}F(x) = R \ne \rho_L .
\end{equation}
On the left boundary, we have
\begin{equation}
\label{left_density}
F(0) = 0 .
\end{equation}
This boundary condition  can be understood as follows. A site located  near the left boundary can be occupied only if the particle leaving the special site at $L$  lands on it. This is a rare event (of probability $\sim 1/L$)  and therefore the microscopic density $\rho_i$ ($i=1,2,\dots$) vanishes in the limit $L \to \infty$.
This observation is also supported by numerical simulations.

Integrating \eqref{Euler} over the interval $0<x<1$ and using \eqref{left_density} we obtain 
\begin{equation}
\label{betarhoL}
\beta \rho_L =  R(1-R)  .
\end{equation}
This is the current of flying particles from the site $L$.
Inserting \eqref{betarhoL} back to \eqref{Euler} we arrive at the  equation
\begin{equation}
\label{Euler_main}
\frac{d}{dx}\,F(1-F) = \frac{R(1-R)}{v}\,(1-F) ,
\end{equation}
which will be  our chief governing differential equation.
Note that $\beta$ does not appear in this equation. 

Equation \eqref{Euler_main} is akin to the one obtained by  Parmeggiani, Franosh and Frey 
(PFF) in their study of the TASEP with open boundaries and Langmuir kinetics (adsorption onto empty sites and desorption from occupied sites) in the bulk \cite{PFF1,PFF2} where the total number of particles is not conserved. We emphasize that in our system  there is neither injection nor  extraction of particles (see  also \cite{EJS,PRWKS,MK,JWKHHW} for related models). A more precise discussion of the relation  between our system and the PFF model will be given
in Sec.~\ref{sec:selection}.

In the following two sections, we shall study the possible solutions to Eq.~\eqref{Euler_main}.
The selection of the correct  density profile $F(x)$ will be achieved by going 
beyond the simple hydrodynamic limit \eqref{Euler_main}.
We shall discuss this in Sec.~\ref{sec:selection},
where the parameter $\beta$ appears again.

\section{Smooth Density Profile}
\label{sec:smooth}

\begin{figure}
\includegraphics[width=9cm]{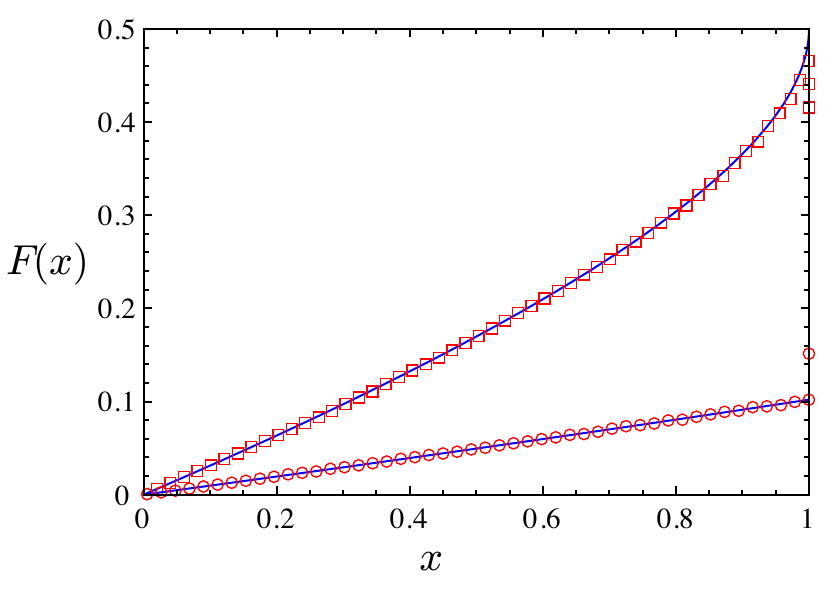}
\caption{Smooth profiles.
The top curve is the critical profile i.e. $v=v_c$
and the lower curve is a sub-critical profile with $v=0.95$. 
They are realized by Monte Carlo simulations.
The plot markers $\square$  and $\circ$
correspond to simulation results with 
parameters $(v,\beta)=(v_c,0.6)$ and 
$(v,\beta)=(0.95,0.6)$, respectively.
We have set $L=10^5$ sites 
and took an average over time steps 
 $10^{11}\le T\le 2\times 10^{11} $. 
}
\label{fig:smooth}
\end{figure}

We first suppose that the density profile is smooth in the bulk. Solving \eqref{Euler_main}
subject to \eqref{left_density} we obtain
\begin{equation}
\label{eq:smooth-profile}
2F + \ln(1-F) = \frac{R(1-R)}{v}\,x  .
\end{equation}
This is valid for all $0 <  x <  1$. 
Combining \eqref{eq:smooth-profile} with \eqref{lim_right_density} we get
\begin{equation}
\label{R}
  2R+\ln(1-R) =  \frac{R(1-R)}{v}\, .
\end{equation}
It is straightforward to check that Eq.~\eqref{R}
may only be satisfied for 
\begin{equation}
\label{eq:v>=vc}
   v\geq v_c = \frac{1}{4(1-\ln 2)} = 0.8147228\dots ,
\end{equation}
where $v=v_c$ corresponds to $R(v_c)=\frac{1}{2}$. Conversely, for any $v \geq v_c$, there exists a unique smooth density profile obtained as follows. We let $R=R(v)$ be the unique solution of \eqref{R} with $R\in (0,\tfrac{1}{2}]$.
 Then, since the function $2F + \ln(1-F)$ is increasing when
 $F\in(0,\tfrac{1}{2})$, Eq.~\eqref{eq:smooth-profile}
  determines consistently $F(x)$ on the entire spatial range $0<x<1$. (Note that, since $ 2F + \ln(1-F)$ decreases when $F\in(\tfrac{1}{2},1)$, a smooth solution cannot exceed the density $\tfrac{1}{2}$.) When $v=v_c$, we have
\begin{equation}
\label{eq:crit}
2F + \ln(1-F) = x(1-\ln 2) .
\end{equation} 
In Fig. \ref{fig:smooth}, 
we plot the critical profile \eqref{eq:crit} and a sub-critical profile  
\eqref{eq:smooth-profile} arising at a certain $v>v_c$.
Note that the critical density profile \eqref{eq:crit}
 has a singularity near $x=1$, viz.
$\tfrac{1}{2} - F(x)\sim \sqrt{1-x}$.
The analytical expressions \eqref{eq:smooth-profile}  and \eqref{eq:crit} are confirmed by 
Monte Carlo simulations (see further discussion in Sec.~\ref{sec:selection}).

We performed simulations using the following algorithm. 
In each time step $T$,  we randomly choose one site $j$ among $L$ sites.
If site $j\le L-1$ is chosen and $(\tau_j,\tau_{j+1})=(1,0)$,
we move the particle at site $j$   as $(\tau_j,\tau_{j+1})=(0,1)$.
If site $L$ is chosen  and $\tau_L=1$, we make 
the particle at site $L$ fly with probability $\beta$ 
to a randomly chosen empty site.  (When $\beta>1$ we have to modify the strategy.
If site $j\le L-1$ is chosen and  $(\tau_j,\tau_{j+1})=(1,0)$,
we move the particle at site $j$  with probability $p$.
If site $L$ is chosen  and $\tau_L=1$, we make 
the particle at site $L$ fly with probability $p\beta$.
We need to chose $p$ such that $p\beta\le 1$.) 

\section{Shock Wave Formation}
\label{sec:shock}

When the fraction of vacant sites is below the critical value,
$v<v_c$, the density profile is no longer described by
Eq.~\eqref{eq:smooth-profile} on the entire spatial range $0<x<1$. In
principle, there may be several patches where the density increases
but remains smaller than $\tfrac{1}{2}$, alternating with patches
where the density decreases and remains higher than $\tfrac{1}{2}$.
(Note that, in the argument of this section, we do not need to impose
the restriction $v<v_c$.)

The only consistent stationary arrangement has precisely two
such patches, a low density patch on the left and a high density patch 
on the right. To appreciate this assertion we recall that in asymmetric exclusion processes shock
waves result from jumps from low to high density, while jumps from
high to low density yield rarefaction waves which are non-stationary (the width of the region covered by a rarefaction wave increases linearly in time). The densities on both sides of a
stationary shock wave at position $s$ are related by current
conservation (Rankine-Hugoniot condition), which for the TASEP 
implies that the densities before and after the shock sum up to 1. We denote the densities before 
and after the shock by $r$ and $1-r$, respectively; from the above discussion we expect that $r\leq \tfrac{1}{2}$. We
confirm this bound by Monte Carlo simulations.

In the region before the shock, we have, by \eqref{Euler_main} and
\eqref{left_density},
\begin{equation}
\label{profile_before}
2F + \ln(1-F) = \frac{R(1-R)}{v}\,x \quad (0<x<s)
\end{equation}
The location of the shock $s$ and the density $r$
just before the shock are related via
\begin{equation}
\label{rvs}
2r + \ln(1-r) = \frac{R(1-R)}{v}\,s
\end{equation}

In the complementary spatial interval $(s,1)$, the solution to  Eq.~\eqref{Euler_main} subject to the boundary condition $\lim_{x\downarrow s}  F(x) = 1-r$ is 
\begin{equation}
\label{profile_after}
2(F-1+r) + \ln\frac{1-F}{r} = \frac{R(1-R)}{v}\,(x-s). 
\end{equation}
Combining this solution with the boundary condition \eqref{lim_right_density} we obtain
\begin{equation}
\label{Rvs}
2(R-1+r) + \ln\frac{1-R}{r} = \frac{R(1-R)}{v}\,(1-s) . 
\end{equation}

We thus have two equations, \eqref{rvs} and \eqref{Rvs},  for three
variables $s,r,R$. We must also keep in mind the bounds on these
variables (following from the above discussion):
\begin{equation}
\label{bounds}
0\leq s\leq 1, \quad 0\leq r\leq \tfrac{1}{2}
\leq R \leq 1-r.
\end{equation}

\begin{figure}
  \includegraphics[width=7.5cm]{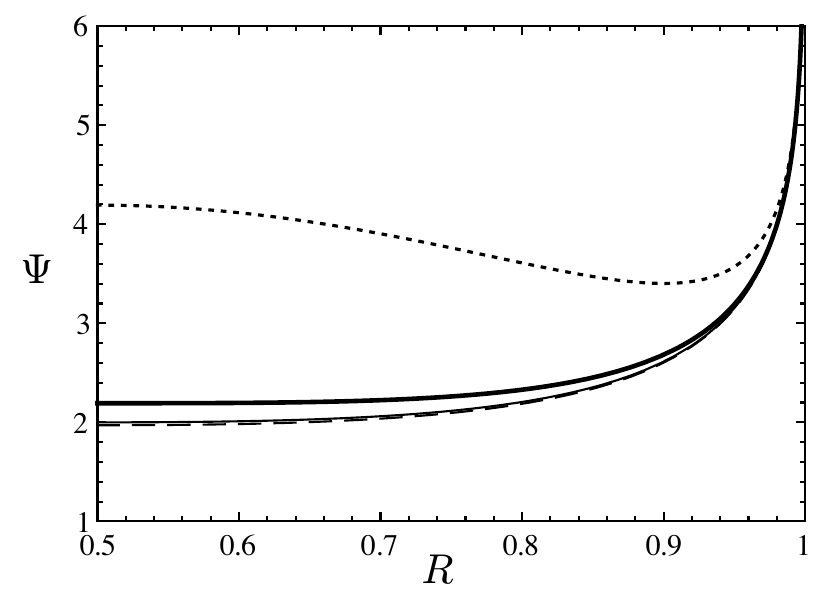} 
  \includegraphics[width=7.5cm]{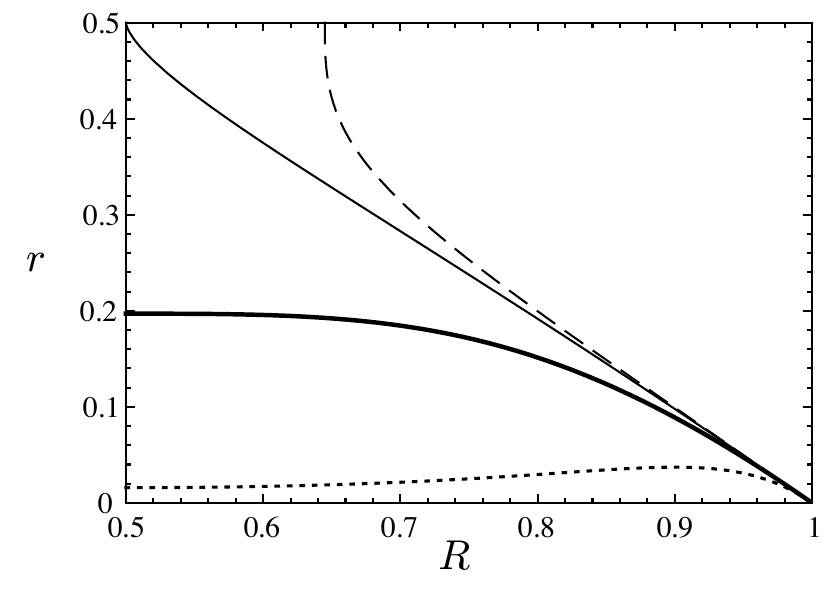} 
  \includegraphics[width=7.5cm]{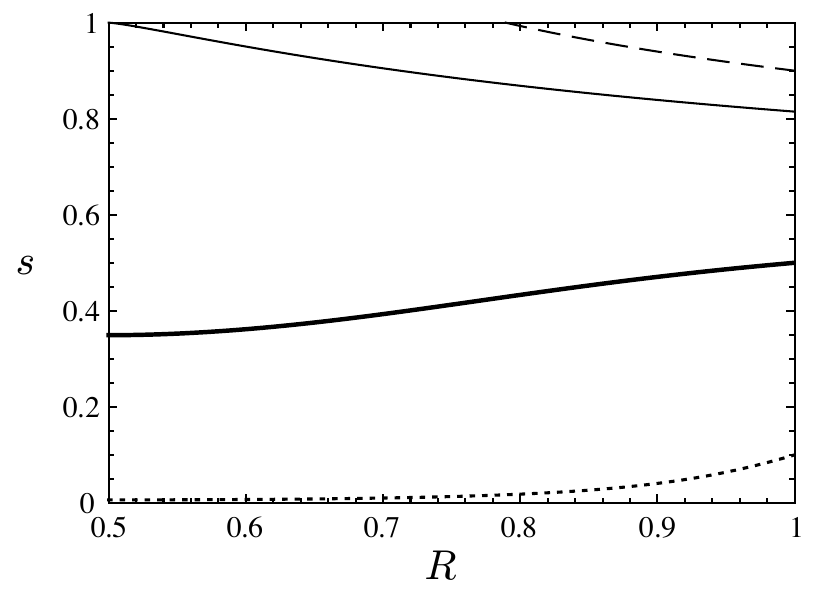} 
\caption{
The right-hand side of Eq.~\eqref{eq:rR} (top),
the density just before the shock $r$ (middle) and 
the position of the shock $s$ (bottom)
are determined by $R$. The fraction  of the vacant sites is chosen as 
the following four values: $v=$0.1 (dotted), 0.5 (bold), $v_c$, and  0.9 (dashed). 
}
\label{fig:R-vs-three}
\end{figure}
 Adding \eqref{rvs}
and \eqref{Rvs} we obtain
\begin{equation}
\label{eq:rR}
4r+ \ln\frac{1-r}{r} = \Psi(R,v) ,
\end{equation}
where we have used the shorthand notation
\begin{equation}
\Psi(R,v) = \frac{R(1-R)}{v} + 2(1-R) - \ln(1-R) .
\end{equation}
The variable $s$ can then be recovered from either \eqref{rvs} or \eqref{Rvs}.

The left-hand side of  Eq.~\eqref{eq:rR} is a monotonically decreasing
function of $r$, thus $r$ is determined according to a given value of
$R$. Then $s$ is fixed by inserting the values of $R$ and $r$ into
\eqref{rvs} or \eqref{Rvs}. On the other hand, the right-hand side
$\Psi$ of Eq.~\eqref{eq:rR} is not always monotonic. For
$v\geq\frac{1}{2}$, $\Psi$ monotonically increases, where the maximal
$r$ is achieved by $R=\tfrac{1}{2}$. For   $v< \tfrac{1}{2}$,  $\Psi$
decreases on the interval $\tfrac{1}{2}<R<1-v$ and increases on the
interval $1-v<R<1$. Thus $r$ takes its maximal value at $R=1-v$. For
example, the dotted lines for $v=0.1$ in the top and middle graphs of
Fig.~\ref{fig:R-vs-three} show $\Psi$ takes the minimal value and $r$
takes the maximal value at $R=1-v=0.9$.  In both cases
($v\geq\frac{1}{2}$ and $v<\frac{1}{2}$), the infimum $r=0$ is
achieved by the limit $R\to 1$, where the shock position $s$
approaches $v$.  In the middle and bottom graphs of
Fig.~\ref{fig:R-vs-three}, we notice  that  the dashed curves for
$v=0.9$ reach $r=0.5$ and $s=1$, respectively.  There exits a minimum
$R_m$ such that $r$ and $s$ satisfy the condition \eqref{bounds}: for
$v<\frac{1}{2}$, $R_m$ is $\frac{1}{2}$, and for $v\geq\frac{1}{2}$,
$R_m$ is the solution to  
\begin{equation}
  2(1-R_m) + \ln R_m = \frac{ R_m(1-R_m)   }{v}.
\end{equation}

\begin{figure}
\includegraphics[width=9cm]{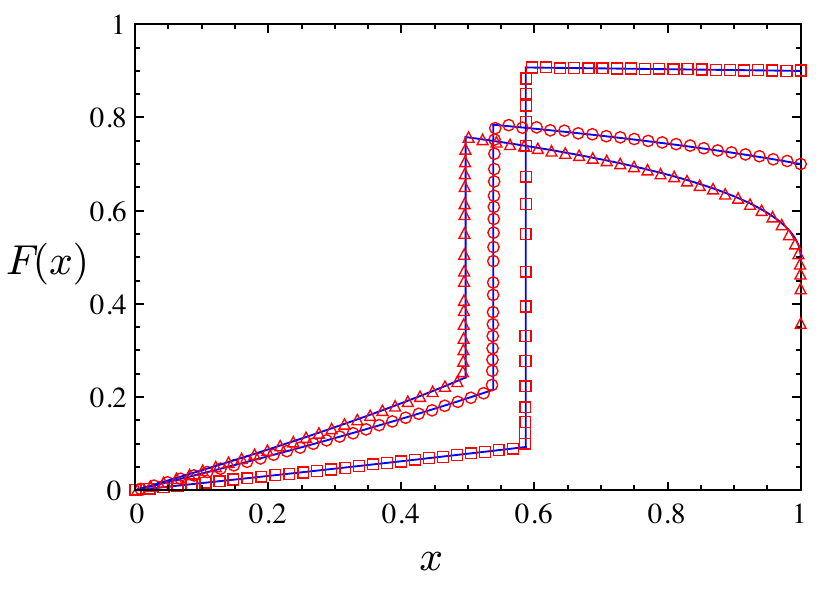} 
\caption{
The density profiles [Eqs. \eqref{profile_before} and
 \eqref{profile_after}] 
 with a shock arising at $v=0.6$ and $R=0.9, 0.7, 0.5$.
They are realized by simulations 
with $\beta=0.1 (\square), ~0.3(\circ), ~0.7 (\triangle)$, respectively,
where $L=10^5$ and the averages over time steps 
 $10^{11} \le T\le 2\times 10^{11}$ were  taken. 
}
\label{fig:shock}
\end{figure}

We have seen that the density $r$  before the shock and 
the shock position $s$ are specified by $R$.
In other word, once one gives a value for $R\in [R_m, 1]$, 
the shape of the density profile is uniquely fixed.
In this sense shock profiles 
construct a one-parameter ($R$) family of solutions.
Figure \ref{fig:shock} provides shock profiles with  $v=0.6$ fixed 
and $R$ varied. These profiles arise in Monte Carlo simulations as 
we will demonstrate in Sec.~\ref{sec:selection}.

The results of the previous section  
tell us that the shape of the smooth profile is uniquely determined when $v\geq v_c$.
However, in this region, one can also construct shock profiles by patching two smooth solutions.
For $v<v_c$, we have shown that the smooth profile {\it ceases to exist} and
 the solution ought to have a shock. Shock profiles are characterized by the parameter $R$.
A selection mechanism has to be found to explain
when and why the system ``prefers'' a continuous or discontinuous profile,
and how the parameter $R$ is determined by $(v,\beta)$.

We have argued that the correct solution either has no shock or a single shock.
Multiple shocks may arise in different models:
for instance, in a system with open boundaries and Langmuir kinetics in the bulk
solutions with two shocks have been detected for a certain exclusion process with interaction between neighboring particles \cite{PRWKS}.

\section{The selection mechanism}\label{sec:selection}

We have obtained  a smooth solution and  
a one-parameter family of shock solutions of the Euler equation \eqref{Euler_main}.
In certain range of filling fractions, we also have a competition between smooth and discontinuous profiles.

First, we remark that the analysis carried out in the previous sections was based on the differential equation~\eqref{Euler_main} 
with the boundary condition~(\ref{left_density});
 in particular, the flying rate $\beta$ has played no role in discussing the two types  of solutions. Therefore, we must incorporate in our analysis the boundary condition in the vicinity of the site $L$ by taking into
account the sum rule~(\ref{betarhoL}).

To determine the phase diagram of the system,
we shall keep  a sub-leading viscosity  term to 
the mean field equation,  construct the  right boundary layer
and apply a stability  analysis to it.
The stationary hydrodynamic equation with viscosity
is given by (see Appendix for a derivation):
\begin{equation}
\frac{1}{2L} \frac{d^2}{dx^2} F
 -\frac{d}{dx} F(1-F) + \frac{\beta}{v} \rho_L(1-F) =0 .
\label{eq:avecviscosite}
\end{equation}
We are interested in what happens in the neighborhood of $x =1$. Defining the local variable $x = 1-\frac{\xi}{2L}$ and writing $f(\xi) = F(1-\frac{\xi}{2L})$, we get
\begin{equation}
   \frac{d^2}{d\xi^2} f +\frac{d}{d\xi} f (1-f) = 0,
\end{equation}
as $L\to\infty$ (after neglecting terms of order $L^{-1}$). Integrating this equation
with the condition  $f(+\infty)=R$, we obtain 
\begin{align}
   \frac{d f}{d\xi}   =  R(1-R) - f(1 -f) = (R-f)(1-R-f) \,.
 \label{eq:Layer}
\end{align}
This is the boundary layer equation that has to be solved subject to the initial condition 
$f(0) = \rho_L = \frac{R(1 -R)}{\beta}$.
We note that \eqref{eq:Layer} has  two fixed points $R$ and $1 -R$.
Setting $f=R+h$ (with $h \ll 1$), we have 
\begin{equation}
\label{dhdxi=}
   \frac{d}{d\xi} h  = (2R-1) h .
\end{equation}
Therefore, if $R < 1/2$,  the fixed point $f = R$ is stable and $f=1 -R$ is unstable.
If $R >  1/2$, the fixed point  $f = R$ is unstable and  $ f=1 -R $ is stable. We also recall 
(Secs.~\ref{sec:smooth} and \ref{sec:shock})  that smooth profiles
are characterized by the fact that $R \le 1/2$,  whereas for shock solutions we always have $R \ge 1/2$.

To determine the phase diagram, consider first the situation when a shock profile is selected ($R\ge \frac{1}{2}$). If  $R >  \frac{1}{2}$, then, according to  Eq.~\eqref{dhdxi=},
$f = R$  is unstable and no boundary layer matching 
$f(0) = \rho_L = \frac{R(1 -R)}{\beta}$  with $R$ can exist. Only one possibility remains: 
$R=  \rho_L = \frac{R(1 -R)}{\beta}$. This implies 
\begin{align}
 R = 1 - \beta \,
\end{align}
which is realized only when  $\beta < \frac{1}{2}$
because of the assumption $R >  \frac{1}{2}$.
 
If the bulk value is $R =  \frac{1}{2}$, it can connected to the boundary value 
$\rho_L  = \frac{1}{4 \beta}$ through a boundary layer.  Using  Eq.~\eqref{eq:Layer} we observe that
$f(\xi)$ is  increasing with $\xi$ i.e. $\rho_L < R = 1/2$ which in turn implies  $\beta > \frac{1}{2}$; thus,  for
$\beta$ greater  than 1/2, the selected  shock profile is the one with  $R =  \frac{1}{2}$.  

To summarize, when the selected profile has a shock, the parameter $R$ is given by 
\begin{align}
\label{R=}
 R=
\begin{cases}
  \frac{1}{2} & (\beta>\frac{1}{2})  , \\
    1-\beta & (\beta\le\frac{1}{2}) ,
\end{cases}
\end{align}
where the true density of the rightmost site is given by
\begin{equation}
\rho_L =
\begin{cases}
  \frac{1}{4\beta} \neq R  \quad  \,\, \,\, \,
  \hbox{ {\small [Boundary Layer]}}& (\beta>\frac{1}{2}), \\
   1-\beta =R     \quad \hbox{{\small  [No Boundary Layer]}}  &
     (\beta\le\frac{1}{2}) .
\end{cases}
\end{equation}

\begin{figure}
 \includegraphics[width=6cm]{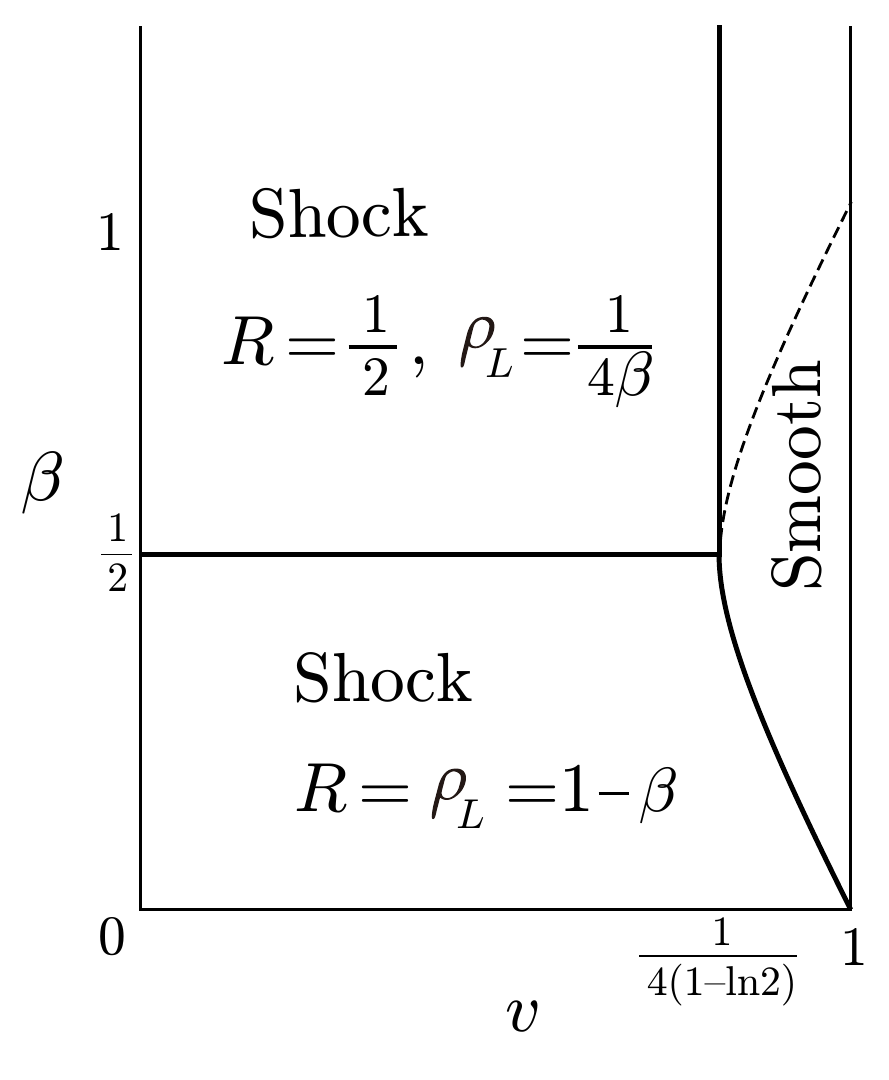} 
\caption{The phase diagram of the model. The full lines represent the boundaries between the three
phases of the model: a phase with smooth density profile and two domains where a shock appears.
The two shock phases differ at the  right boundary: When $R \neq \rho_L,$  a boundary layer is formed; when $R = \rho_L = 1 - \beta$, the solution of the inviscid hydrodynamic equation provides an excellent description of the density profile of the entire system.}
\label{fig:phase-diagram}
\end{figure}

To understand when the shock profile with parameter $R$ given by \eqref{R=} can be 
selected we set the shock position $s=1$ and we observe that a shock can be found only when $v<v_c (\beta)$, with  
\begin{equation}
 v_c (\beta) = 
\begin{cases}
  v_c = \frac{1}{4(1-\ln 2)}  & (\beta>\frac{1}{2})  , \\
   \frac{\beta(1-\beta)}{2\beta+\ln(1-\beta) } & (\beta\le\frac{1}{2}) ,
\end{cases}
\end{equation}
which is drawn in Fig. \ref{fig:phase-diagram}.

By contraposition, only  the smooth profile is selected  for $v\ge v_c(\beta)$.
As we have shown in Sec.~\ref{sec:smooth}, a smooth profile cannot exist for  $v < v_c$.
Hence,  at this stage, we have not yet determined which profile is selected 
in the region  $v_c<v< v_c(\beta),$ and $ \beta<\frac{1}{2}$.
We now  show that a  smooth profile is not allowed for $v< v_c(\beta),\ \beta<\frac{1}{2}$.
Let us assume, by  {\it reductio ad absurdum,} that a  smooth profile exists.
The parameter  $R < 1/2$ is a stable fixed point of  ~\eqref{eq:Layer}, and must be connected to $ \rho_L = \frac{R(1 -R)}{\beta}$ through a boundary layer satisfying  Eq.~\eqref{eq:Layer}. But  Eq.~\eqref{eq:Layer} has {\it two} fixed points $R$ and $1 - R$. Because this is a first order equation, no solution can
cross a fixed point, i.e., the boundary layer that connects $R$ and $\rho_L$ can not pass through the value  $1-R$. This implies $\rho_L \le 1-R$, i.e., $R \le \beta$. This condition gives $v\ge v_c(\beta)$.
In other words, a smooth profile cannot exit when $v<v_c(\beta)$.

\begin{figure}
 \includegraphics[width=9cm]{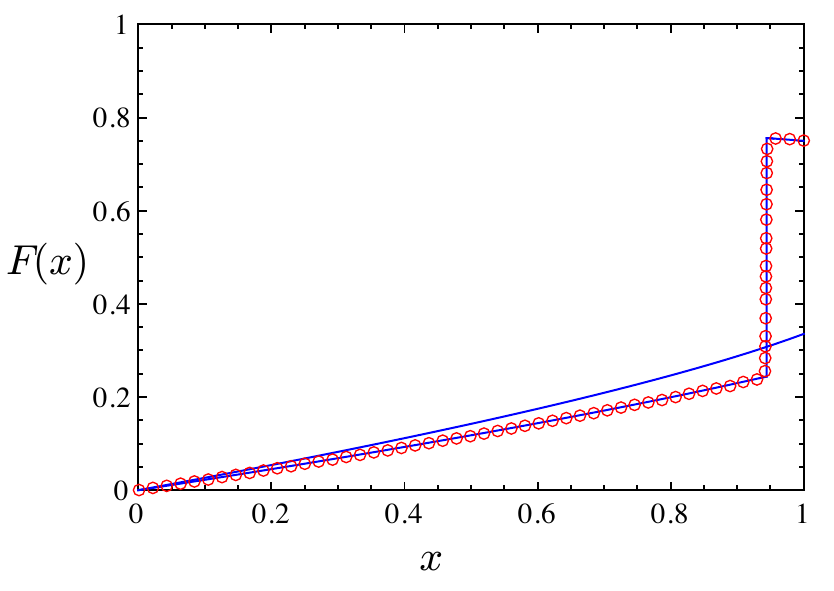} 
\caption{
A smooth profile and a shock profile
are  solutions to Eq. \eqref{Euler_main} with $v=0.85$ (solid lines).
For the shock profiled, we have set the parameter $R=0.75$.
The plot markers $\circ$ correspond to 
simulation result with $(v,\beta) = (0.85,0.25)$,
where $L=10^5$ and 
 the average over time steps 
 $ 10^{11} \le T\le 2\times 10^{11}$ was taken. 
We observe from  the simulation that the  shock profile is selected 
with  $R=1-\beta$.}
\label{fig:smooth-shock}
\end{figure}

The full phase diagram of the system is represented in Fig.~\ref{fig:phase-diagram}.
 There are  three phases: a phase with smooth density profile and two domains where a shock appears. The shock phases differ with respect to the boundary layer. 
When $\beta > 1/2$, a boundary layer is formed; for $\beta < 1/2$, the profile is smooth 
in the vicinity of $x =1$ (see  Fig.~\ref{fig:smooth-shock}). The dotted line 
in the smooth phase is  $v=\frac{\beta(1-\beta)}{2(1-\beta) + \ln\beta}$
which  corresponds to the accidental case 
where there is no boundary layer $  \rho_L = R= 1-\beta $. Hence, generically,  in the smooth phase
a boundary layer always exists.

From this full phase  diagram, we can deduce  the phase diagram of the original problem in which we were interested.
 There, the particle at site $L$  attempts  to hop 
to a randomly chosen site with rate unity and the jump  is  accepted if the
target site is  empty. This 
 corresponds to the diagonal line
 $\beta = v$ in the phase diagram in Fig.~\ref{fig:phase-diagram}.
We observe that there are also three phases for this special problem:
for $v \ge  v_c = [4(1-\ln 2)]^{-1}=0.8147228\ldots$, the profile is smooth;
for $1/2 < v  <  v_c$,  the profile is of shock type  with  $R = 1/2$ and  there exists a boundary layer;
for  $0 < v \le 1/2$,  the profile is of shock type  with    $R = 1 - v$ and there is no boundary layer. 
We notice that $R$ is chosen so that $r$ is maximized. 

\begin{figure}
 \includegraphics[width=9cm]{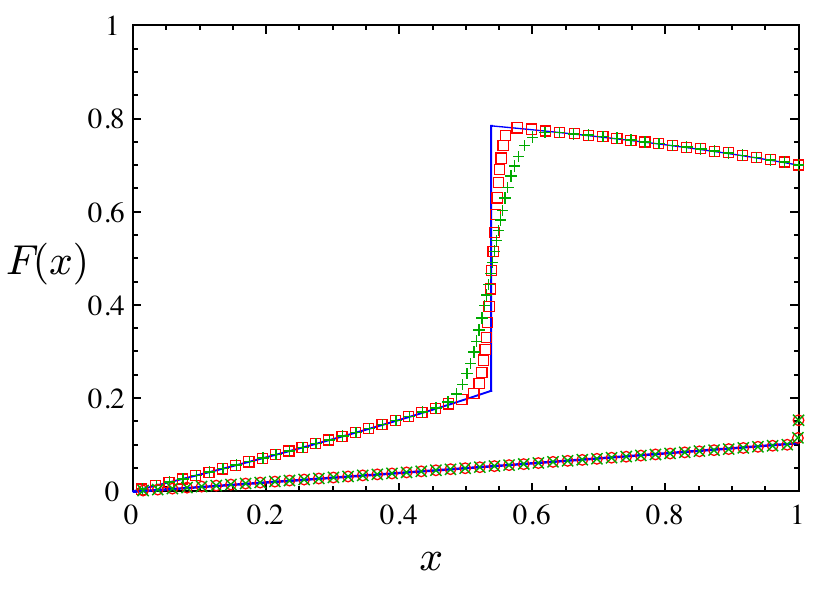} 
\caption{
Comparison with a special case of the PFF model.
The parameters of our model are chosen as 
$(\beta,v)=(0.6, 0.95)$ ($\circ$, smooth profile)
and  $(\beta,v)=(0.6, 0.3)$ ($\square$, shock profile),
 corresponding to the PFF model
  with $(\beta,\Omega_a)=(0.6,0.09645\dots)$ ($\times$)
  and $(0.6,0.8)$ ($+$), respectively.
In both models we have set $L=10^3$
(and we took an average over time steps 
 $5\times 10^8 \le T\le 5\times 10^9 $),
but we observe difference around the shock position.
}
\label{fig:comp-PFF}
\end{figure}

Finally we mention a relation between our system
and the PFF model \cite{PFF1, PFF2}.  
The PFF model is defined on an  finite open lattice by the following rules:
\begin{eqnarray*}
   10 &\to& 01  \quad  \text{at rate } 1 \, , \\ 
   0 &\to& 1    \quad \;  \text{ at rate } \Omega_a/L \, ,\\
   1 &\to& 0    \quad \;   \text{ at rate } \Omega_d/L \, ,
\end{eqnarray*}
in the bulk, and 
\begin{eqnarray*}
   0 &\to&  1   \ \quad   \text{ at rate }\alpha \, ,\\ 
    1 &\to&  0   \  \quad  \text{ at rate }\beta   \, ,
\end{eqnarray*}
at the left and right boundaries, respectively.
The rates $\alpha$ and $\beta$
are related to the densities of the left and right reservoir, respectively.
We note  that the hydrodynamic equation that describes 
the macroscopic density profiles of our model can be formally  mapped into
the one corresponding to the PFF model by setting 
\begin{align}
 \alpha=\Omega_d=0, \  \Omega_a=  \frac{R(1-R)}{v} .
\end{align}
This ensures that at the hydrodynamic level
 [Eq.~\eqref{Euler_main}] the models coincide. 
However, we emphasize that the factor $\frac{R(1-R)}{v}$ is not determined by external conditions but
 rather by $R=F(1)$ and by Eq.~\eqref{Euler_main}.
Besides, the system we consider is conservative and therefore the mapping to the PFF model
can only be approximate at the mean-field level.
This fact can be observed in 
Monte Carlo simulations which  show  significant differences between the shock  profiles  of the two models for a given  system size, see Fig.~\ref{fig:comp-PFF}. Thus, 
it  would be  therefore of interest to explore further the connection between our model  and the PFF model  \cite{PFF1, PFF2}. We emphasize that whereas the number of particles
is strictly  conserved in the process we study,  the PFF model  exchanges particles with its surroundings.
The correspondence can not be perfect and it is also that  for non-equilibrium systems,
 where the equivalence of ensembles (such as canonical vs grand-canonical)
is a subtle  issue, especially when shocks or phase-separation are involved \cite{AHR,RSS}. In particular, 
fluctuations  can be significantly  different.

\section{Concluding remarks}
\label{sec:concluding-remarks}

We studied an exclusion process with a special site from which non-local hoppings are allowed.
We showed that a shock is formed when the average density exceeds the critical value.   We established a phase diagram of the model by using a hydrodynamic approach, a boundary layer analysis, and an  asymptotic matching. It would be interesting to devise a physical principle, perhaps reminiscent to the extremal current principle suggested by Krug \cite{K}  (see also
\cite{KSKS,PS,HKPS,DME}), that would lead to the same phase diagram. Overall our model provides another manifestation of a remarkable phenomenon, namely that a single defect can have a drastic effect on non-equilibrium steady states.

Although our analysis relies on a mean-field approximation, the results appear asymptotically exact and they are in excellent agreement with Monte Carlo simulations. We performed simulations 
for large sizes (up to $L = 10^5$) at over 100 points $(v,\beta)$ in the phase diagram.
Our model demonstrates again that localized defects in far-from-equilibrium systems can produce global effects that drastically alter the phenomenology of a model. Systems far from equilibrium
are therefore very sensitive to boundary conditions and to defects. 

Models with periodic  and open boundaries have been studied extensively, yet little is known about inhomogeneous systems \cite{JL1,Gunter,JL2,CLST}.  We believe that the interplay of impurities and currents can lead to rich and unexpected behavior (note that even in thermal equilibrium the presence of a magnetic impurity can result in  the highly non-trivial Kondo effect). In biological applications, inhomogeneities are unavoidable, and various semi-realistic inhomogeneous extensions of the TASEP have been analyzed, see e.g. \cite{CL,DSZ,BRGT}.  It would be very interesting to find
and to analyze exactly solvable extensions of the TASEP that are both far from equilibrium and non-homogeneous.

\begin{acknowledgments}
We thank C\'ecile Appert-Rolland, Arvind Ayyer, Thierry Bodineau and Tridib Sadhu
for helpful discussions and suggestions.
\end{acknowledgments}

\appendix
\section{Derivation of the hydrodynamic equations}

Here we explain how the hydrodynamic limit \eqref{Euler} and the
viscosity correction can be obtained starting from the microscopic
dynamics. This is a standard  procedure  (see e.g. \cite{K,BA})
which we recall for the sake of completeness.  

Exact dynamical equations can be written for the
expectation value $\rho_j=\langle \tau_j  \rangle$  of $\tau_j$
(i.e. the density of site $j$) at time $t$ as 
\begin{align}
   \begin{split}
 \label{eq:bulk}
\frac{d \rho_j}{ dt}  = &   
\langle \tau_{j-1}( 1 - \tau_j) \rangle 
 - \langle \tau_{j}( 1 - \tau_{j+1}) \rangle \\
  & +\frac{\beta}{L v} \langle \tau_{L}( 1 - \tau_j) \rangle 
   \quad (j=2,3,\dots,L-1)\, , 
\end{split}
\\
\label{eq:drho1dt=}
    \frac{d \rho_1}{ dt}   =&
 \frac{\beta}{L v}  \langle \tau_{L}( 1 - \tau_1) \rangle 
 - \langle \tau_{1}( 1 - \tau_{2}) \rangle \, , \\
    \frac{d \rho_L}{ dt}  =   &
 \langle \tau_{L-1}( 1 -\tau_L) \rangle - \beta \rho_L \, .
\end{align} 
As usual, Eqs.~\eqref{eq:bulk} are part of a hierarchy that couples
correlations of a given order  to higher order correlations. Here,
this hierarchy will be closed  by assuming that in the hydrodynamic
limit $L \to \infty$, the local density  profile $F(x)$ defined in
(\ref{HydrProf}) satisfies a mean-field equation. This may sound like a 
very bold assumption but it is known to lead to sound results for exclusion processes.

However, we shall go beyond the Eulerian inviscid treatment by keeping track of viscosity terms that scale as $L^{-1}$. These higher-order contributions will have an important physical role within the boundary layers.

 In the bulk, we have
\begin{equation}
\frac{ \partial F}{ \partial t} = \frac{ 1} { 2 L^2} 
\frac{ \partial^2  F }{ \partial x^2}  -
 \frac{1}{L} \frac{\partial }{ \partial x}\, F(1 - F) 
 +  \frac{\beta }{L v} \rho_L(1-F)
\end{equation}
where $x = j/L$ and $ \rho_L \equiv \langle \tau_L \rangle$.  Rescaling the time as $ t \to L t$ 
and going to the stationary limit, we obtain 
\begin{equation}
 \frac{1}{2L} \frac{ d^2 F }{d x^2}
  -  \frac{d}{d x}\, F(1 - F)
 +  \frac{\beta}{v} \rho_L(1-F) = 0 \,.
 \label{ViscousHydro}
\end{equation}
The viscosity term is a singular perturbation that can be neglected in the regions where $F$ varies smoothly (i.e.,  outside the shock regions and the boundary layers); Eq.~(\ref{Euler}) is then
recovered. Near $x=1$, a  boundary layer may exist and this fact  plays a  crucial role in selecting the {\it global} density profile: this is precisely what is studied in Sec.~\ref{sec:selection}. Near $x=0$,
we observe $F(0)=0$ and there is no boundary layer. This is supported by the fact the 
stationary current near $x=0$ vanishes  in the limit $L\to\infty$.
For example, from Eq.~\eqref{eq:drho1dt=} one indeed finds that the current from site 1 to site 2 vanishes as the system size diverges, viz.
$\langle \tau_{1}( 1 - \tau_{2}) \rangle \sim L^{-1}\to 0$
when $L\to\infty$.


\begin{thebibliography}{99}

\bibitem{MGP}
     C. MacDonald, J. Gibbs, and A. Pipkin, Biopolymers {\bf 6}, 1 (1968); 
     C. MacDonald and A. Pipkin, Biopolymers {\bf 7}, 707 (1969). 

\bibitem{Sp}   
     H. Spohn, {\it Large Scale Dynamics of Interacting Particles} 
     (New York: Springer-Verlag, 1991).    


\bibitem{K}   J. Krug, Phys. Rev. Lett.  {\bf 67}, 1882 (1991).


\bibitem{HZ} T. Halpin-Healy and Y.-C. Zhang, Phys. Rep. {\bf 254}, 215 (1995).

\bibitem{Sch1} 
    G. M. Sch\"utz, Exactly Solvable Models for Many-Body Systems
    Far From Equilibrium, in {\it Phase Transitions and Critical Phenomena},
    Vol.\ 19, eds.\ C. Domb and J. L. Lebowitz (Academic Press, London, 2000).

\bibitem{BA} 
    R. A. Blythe and M. R. Evans, J. Phys.\ A {\bf 40}, R333 (2007).
          
\bibitem{D1}
     B. Derrida, J. Stat. Mech. P07023 (2007).

\bibitem{DEHP}     
      B. Derrida, M. R. Evans, V. Hakim and V. Pasquier, 
      J. Phys. A: Math. Gen.  {\bf 26}, 1493 (1993).

\bibitem{SD} 
      G. M. Sch\"utz and E. Domany, J. Stat. Phys.  {\bf 72}, 277 (1993).

\bibitem{DEMu}    
      B. Derrida, M. R. Evans, and D. Mukamel, J.\ Phys. A {\bf 26}, 4911 (1993).

\bibitem{DEMa}    
      B.~Derrida, M.~R.~Evans, and K.~Mallick, J.\ Stat.\ Phys. {\bf 79}, 833 (1995).

\bibitem{Sch2}
       G. M. Sch\"utz, J. Stat.\ Phys. {\bf 88}, 427 (1997);
       R. J. Harris, A. R\'akos, and G. Sch\"utz, J. Stat. Mech. P08003 (2005).

\bibitem{PM}
      S. Prolhac and K. Mallick, J. Phys. A {\bf 41}, 175002 (2008); 
      J. Phys. A {\bf 42}, 175001 (2009).

\bibitem{TW} 
      C. Tracy and H. Widom,  J. Stat.\ Phys. {\bf 132}, 291 (2008); 
      Commun.\ Math.\ Phys. {\bf 290}, 129 (2009); J. Stat.\ Phys. {\bf 140}, 619 (2010). 

\bibitem{D2}  B. Derrida, J. Stat. Mech.  P01030 (2011).

\bibitem{GCCR} A. Garai, D. Chowdhury, D. Chowdhury and T. V. Ramakrishnan,
  Phys. Rev. E {\bf 80}, 011908  (2009).

\bibitem{CSN}  D. Chowdhury, A. Schadschneider, and K. Nishinari, 
 Phys. Life Rev. {\bf 2},  318 (2005).

\bibitem{TPPRC}  F. Turci, A.  Parmeggiani, E.  Pitard, M. C.  Romano, 
 and L.  Ciandrini,  Phys. Rev.  E {\bf 87},  012705 (2013).

\bibitem{CSR}
    L. Ciandrini, I. Stansfield, and M. C. Romano, Phys. Rev. E {\bf 81}, 051904 (2010). 

\bibitem{TEK} J. Tailleur, M. R. Evans, and Y. Kafri,
  Phys. Rev. Lett.  {\bf 102}, 118109 (2009).


\bibitem{CLBCJ} O. Camp\`as, C. Leduc, P. Bassereau, J. Casademunt, J.-F. Joanny,
  and J. Prost,  Biophys. J. {\bf 94}, 5009 (2008).

\bibitem{JEK}  D. Johann, C. Erlenk\"amper, and K. Kruse,
   Phys. Rev. Lett.   {\bf 108}, 258103  (2012).

\bibitem{MRF} 
   L. Reese, A. Melbinger, and E. Frey, Biophys. J. {\bf  101}, 2190  (2011);
   A. Melbinger, L. Reese, and E. Frey,  Phys. Rev. Lett. {\bf 108}, 258104  (2012).

\bibitem{EAS} M. Ebbinghaus, C. Appert-Rolland, and  L. Santen,
  Phys. Rev. E {\bf 82},   040901 (2010).

\bibitem{CKZ}  
      T. Chou, K. Mallick and R. K. P. Zia,  Rep. Prog. Phys. {\bf 74}, 116601  (2011).

\bibitem{JL1}
     S. A. Janowsky and J. L. Lebowitz, Phys. Rev. A {\bf 45}, 618 (1992). 

\bibitem{Gunter}
     G. M. Sch\"utz, J. Stat. Phys.  {\bf 71}, 471 (1993). 
     
\bibitem{JL2}
     S. A. Janowsky and J. L. Lebowitz, J. Stat. Phys.  {\bf 77}, 35 (1994). 

\bibitem{CLST}
     O. Costin, J. L. Lebowitz, E. R. Speer, and A. Troiani,
 Inst. Math. Acad. Sin. (N.S.) {\bf 8}, 49 (2013).
%     Acta Math. Scientifica {\bf 8}, 49 (2013). 
%    arXiv:1207.6555. 

\bibitem{EM} M. R. Evans and S. N. Majumdar,
Phys. Rev. Lett. {\bf 106}, 160601 (2011).

\bibitem{PFF1}
      A. Parmeggiani, T. Franosch, and E. Frey, Phys. Rev. Lett. {\bf 90}, 086601 (2003). 
    
\bibitem{PFF2}
      A. Parmeggiani, T. Franosch, and E. Frey, Phys. Rev. E {\bf 70}, 046101 (2004). 
    
\bibitem{EJS}
      M. R. Evans, R. Juh\'{a}sz, and L. Santen,  Phys. Rev.  E  {\bf 68}, 026117 (2003).
      
\bibitem{PRWKS} V. Popkov, A. R\'akos, R. D. Willmann, 
    A. B. Kolomeisky, and  G. M. Sch\"utz,
  Phys. Rev.  E  {\bf 67}, 066117 (2003).


\bibitem{MK} N. Mirin and A. B. Kolomeisky,
  J. Stat. Phys.  {\bf 110}, 811 (2003).


\bibitem{JWKHHW}  R. Jiang, Y.-Q. Wang, A. B. Kolomeisky, W. Huang,
  M.-B. Hu, and Q.-S. Wu,  Phys. Rev.  E  {\bf 87}, 012107 (2013).

\bibitem{AHR}  P. F. Arndt, T. Heinzel  and   V. Rittenberg, J. Phys. A
 {\bf 31}, L45  (1998);   J. Stat. Phys.  {\bf 97},  1  (1999).

\bibitem{RSS} N. Rajewsky, T. Sasamoto, and  E. R.  Speer, 
Physica A: Stat.  Mech. and  Appl. {\bf 279}, 123 (2000).


\bibitem{KSKS}  A. B.  Kolomeisky,  G. M. Sch\"utz,  E. B.  Kolomeisky, and J. P. Straley, J. Phys. A: Math. Gen.  {\bf 31},  6911  (1998).


\bibitem{PS} V. Popkov and  G. M. Sch\"utz, Europhys. Lett. {\bf 48}, 257 (1999).


\bibitem{HKPS} J. S. Hager, J. Krug, V. Popkov and  G. M. Sch\"utz, 
  Phys. Rev.  E  {\bf 63},  056110 (2001).


\bibitem{DME}
    M. Dierl, P. Maass, and M. Einax, Phys. Rev. Lett. {\bf 108}, 060603 (2012);
    M. Dierl,  M. Einax, and P. Maass, Phys. Rev. E {\bf 87},  062126 (2013).

\bibitem{CL}  
      T. Chou and G. Lakatos,  Phys. Rev. Lett. {\bf 93}, 198101  (2004).

\bibitem{DSZ}  
      J. Dong, B. Schmittmann, and R. Zia,  J. Stat. Phys. {\bf 128}, 21  (2007).

\bibitem{BRGT}
    C. A. Brackley, M. C. Romano, C. Grebogi, and M. Thiel, 
    Phys. Rev. Lett. {\bf 105}, 078102 (2010);
     C. A. Brackley, M. C. Romano, and M. Thiel, 
     Phys. Rev. E {\bf 82},  051920 (2010).

\end{thebibliography}
\end{document}